\documentclass[a4paper,superscriptaddress,aps,prb,twocolumn,floatfix,citeautoscript]{revtex4}
\usepackage{natbib}

\newcommand{\vect}[1]{\mathbf{#1}}
\usepackage{graphicx}
\usepackage{color}
\usepackage{amsmath}
\usepackage{amssymb}
\usepackage[english]{babel}

\newcommand{\im}{\text{Im}}
\newcommand{\re}{\text{Re}}
\newcommand{\I}{\text{i}}
\usepackage{bm}
\usepackage{url}
\graphicspath{{figures/}}

\begin{document}
%
% TITLE
% \citenum
\title{Optical spectra of 2D monolayers from time-dependent density functional theory} 
\newcommand{\lpt}{Laboratoire de Physique Th\'eorique, Universit\'e de Toulouse, CNRS, UPS, France}
\newcommand{\lcpq}{Laboratoire de Chimie et Physique Quantiques, Universit\'e de Toulouse, CNRS, UPS, France}
\newcommand{\etsf}{European Theoretical Spectroscopy Facility (ETSF)}
\affiliation{\lpt}
\affiliation{\lcpq}
\affiliation{\etsf}
\author{S. Di Sabatino}
%%\email[]{disabatino@irsamc.ups-tlse.fr}
\affiliation{\lpt}
\affiliation{\lcpq}
\affiliation{\etsf}
\author{J.A. Berger}
\affiliation{\lcpq}
\affiliation{\etsf}
\author{P. Romaniello}
\email{pina.romaniello@irsamc.ups-tlse.fr} 
\affiliation{\lpt}
\affiliation{\etsf}
%
%\pacs{71.10.-w,71.27.+a,31.15.V-,79.60.Bm}

\begin{abstract}
The optical spectra of two-dimensional (2D) periodic systems provide a challenge for time-dependent density-functional theory (TDDFT) because of the large excitonic effects in these materials. In this work we explore how accurately these spectra can be described within a pure Kohn-Sham time-dependent density-functional framework, i.e., a framework in which no theory beyond Kohn-Sham density-functional theory, such as $GW$, is required to correct the Kohn-Sham gap. To achieve this goal we adapted a recent approach we developed for the optical spectra of 3D systems [Cavo, Berger, Romaniello, \textit{Phys.\ Rev.\ B}, \textbf{101}, 115109 (2020)] to those of 2D systems. Our approach relies on the link between the exchange-correlation kernel of TDDFT and the derivative discontinuity of ground-state density-functional theory, which guarantees a correct quasi-particle gap, and on a generalization of the polarization functional [Berger, \textit{Phys.\ Rev.\ Lett.} \textbf{115}, 137402 (2015)], which describes the excitonic effects. We applied our approach to two prototypical 2D monolayers, $h$-BN and MoS$_2$. We find that our protocol gives a qualitative good description of the optical spectrum of $h$-BN, whereas improvements are needed for MoS$_2$ to describe the intensity of the excitonic peaks.
\end{abstract}
\date{\today}
\maketitle
\section{Introduction}
In recent years the research on 2D materials, such as graphene, hexagonal boron nitride ($h$-BN), and  transition-metal dichalcogenides, has grown exponentially in the fields of condensed matter physics, materials science, chemistry, and nanotechnology. Thanks to geometric confinement and reduced dielectric screening these materials exhibit unique features such as strong light-matter interaction and enhanced many-body effects. For example, transition-metal dichalcogenides MX$_2$ (X=S, Se, Te; M=transition metal) represent a particularly interesting class of 2D materials comprising both semiconductors and metals. The prototypical family member MoS$_2$ is well known to undergo a transition from indirect to direct band gap semiconductor  when its thickness is thinned down to a monolayer. 
Furthermore, excitons couple strongly to light and lead to a substantial modification of the optical spectrum both below and above the QP band gap. \cite{PhysRevLett.105.136805,Splendiani,Zhang,Balendhran,Lambrecht,PhysRevB.86.115409,PhysRevB.88.045412,PhysRevLett.111.216805,Thygesen_2017} 

The standard theoretical approach to calculate the optical spectra of 2D materials is by solving the Bethe-Salpeter equation (BSE)~\cite{OnidaReiningRubio} on top of a $GW$~\cite{HedinPR,Aryasetiawan_1998,AulburJonssonWilkins,Reining_2017, Golze_2019} band-structure calculation. This method is usually very accurate since the BSE approach explicitly takes into account both the electron and the hole that make up the excitons.
However, there are also some shortcomings, in particular: 1) the large computational cost of a BSE calculation which, in its standard implementation, scales as $N^6$ (with $N$ the number of electrons) and 2) the difficulty to converge BSE calculations because of the large number of convergence parameters involved. Due to these two shortcomings the $GW$+BSE is, for example, not (yet) suitable for high-throughput screening.

An alternative way to capture the same physics but at a much lower computational cost ($N^3$ or $N^4$ depending on the implementation) and with fewer convergence parameters is time-dependent density functional theory (TDDFT).  
As is well-known, the main difficulty of TDDFT is to find good approximations to the exchange-correlation kernel $f_{\text{xc}}$.
In the specific case of optical spectra, the challenge is two-fold: 1) to capture the two-particle electron-hole interaction within an inherently single-particle picture and 2) to transform the underlying Kohn-Sham (KS) band structure into a quasi-particle band structure.
Almost all recent TDDFT kernels~\cite{bootstrap,Trevisanutto,santiago,Berger,Yang_2015,Terentjev_2018} address the first part but not the second.
Instead, a scissors shift~\cite{Levine_1989} is often used to correct the KS band gap.
The scissors parameter is then obtained either from experiment or from a theory beyond TDDFT such as $GW$ or extended KS theory using hybrid functionals.~\cite{HSE03,HSE06,Matsushita_PRB2011,Galli_PRB2016}

We have recently proposed the Pure functional~\cite{PhysRevB.101.115109} as an approximation to $f_{\text{xc}}$ which describes both the excitonic effects \emph{and} the quasi-particle effects in the optical spectra of solids, i.e., 3D materials.
The spectra we obtained were in good agreement with experimental data.~\cite{PhysRevB.101.115109}
It would be interesting to establish the accuracy of the Pure functional for 2D materials.
However, the Pure functional depends explicitly on the macroscopic dielectric function which is meaningless in 2D materials (see, e.g., \onlinecite{Hueser,Cudazzo,Veniard}).
Therefore, in this work, we propose a modified Pure functional for 2D materials and we assess its accuracy when applied to the calculation of the optical spectra of these materials. We note that very recently Suzuki and Watanabe \cite{Suzuki} have also used TDDFT to calculate the optical spectra of 2D materials.

The paper is organized as follows. In Sec.~\ref{Theory} we give a brief description of the Pure functional derived in Ref.~\onlinecite{PhysRevB.101.115109}. We report the computational details of our calculations in Sec.~\ref{Comput}.
In Sec.~\ref{Results} we show and discuss the results we obtained for tho prototypical 2D materials, namely $h$-BN and MoS$_2$. We finally draw conclusions and perspectives in Sec.~\ref{Conclusions}

\section{Theory\label{Theory}}

The TDDFT problem can be written as a two-point Dyson equation for the polarizability $\chi(\omega)$ according to
\begin{equation}
\chi(\omega)=\chi_{\text{KS}}(\omega)+\chi_{\text{KS}}(\omega)\left[
v_c+f_{\text{xc}}(\omega)
\right]
\chi(\omega),
\end{equation}
where $\chi_{\text{KS}}$ is the Kohn-Sham (KS) polarizability, $v_c$ is the Coulomb potential, and $f_{\text{xc}}$ is the exchange-correlation kernel, 
which is the quantity that has to be approximated in practical applications.
A useful decomposition of the exact $f_{\text{xc}}$ is given by\cite{Sottile_PRL2003}
\begin{align}
f_{\text{xc}}(1,2) &= \underbrace{\chi^{-1}_{\text{KS}}(1,2) - \chi_0^{-1}(1,2)}_{f_{\text{xc}}^{(1)}}
\notag \\ &
\underbrace{-i\int d345 \chi_0^{-1} (1,5)G(5,3) G(4,5) \frac{\delta\Sigma(3,4)}{\delta\rho(2)}}_{f_{\text{xc}}^{(2)}},
\label{Eqn:fxc_exact}
\end{align}
where $\chi_{\text{KS}}$ and $\chi_0=-iGG$ are the Kohn-Sham and independent quasi-particle polarizability, respectively, and $G(1,2)$ and $\Sigma(1,2)$ are 
the one-body Green function and the self-energy, respectively. 
The collective index $(1)=(\mathbf{x},t)=(\mathbf{r},\sigma,t)$ contains the space, spin and time coordinates.
The above decomposition reveals that $f_{\text{xc}}$ has two separate contributions: $f_{\text{xc}}^{(1)}$ which shifts the poles
of the KS polarizabilty to those of the independent quasi-particle polarizability and $f_{\text{xc}}^{(2)}$ which takes
into account the interactions between (quasi-)particles, and, in particular, the electron-hole interaction.

\subsection{The 3D Pure functional}

We have recently shown~\cite{PhysRevB.101.115109} that $f_{\text{xc}}^{(1)}$ can be linked to the derivative discontinuity $\Delta^{\text{dd}}$ of density-functional theory which is defined as the difference between the fundamental gap and the KS gap.
Moreover we have shown that the effect of $f_{\text{xc}}^{(1)}$ on the polarizability can be accounted for by solving the following modified Dyson equation,
\begin{equation}
\chi(\omega)=\chi^{(1)}_{\text{KS}}(\omega)+\chi^{(1)}_{\text{KS}}(\omega)\left[
v_c+f_{\text{xc}}^{(2)}(\omega)
\right]
\chi(\omega),
\end{equation}
where $\chi^{(1)}_{\text{KS}}(\omega)$ is a modified KS response function defined as
\begin{equation}
\chi^{(1)}_{\text{KS}}(\mathbf{x}_1,\mathbf{x}_2,\omega) = \sum_{i,j}
\frac{(f_j - f_i) \phi_i(\mathbf{x}_1)\phi_j(\mathbf{x}_2)\phi^*_j(\mathbf{x}_1)\phi^*_i(\mathbf{x}_2)}{\omega - (\epsilon_i -\epsilon_j) -\text{sgn}(\epsilon_i -\epsilon_j)\Delta^{\text{dd}} +\I\eta}
\label{Eqn:modified-chi},
\end{equation}
where $\phi_i$ is a KS spin orbital, $\epsilon_i$ its energy, $f_i$ its occupation (0 and 1 for unoccupied and occupied orbitals, respectively), and $\eta$ is a positive infinitesimal that ensures causality. 

We will use the GLLB~\cite{GLLB_PRA1995} model to approximate the derivative discontinuity.~\cite{Kuisma_PRB2010,C7CP02123B}
In this model $\Delta^{\text{dd}}$ is given by
%5
\begin{align}
\Delta^{\text{dd}} =& K_{\text{xc}}\sum_{i=1}^N
\left[\sqrt{\epsilon_{\text{CBM}}-\epsilon_i}-\sqrt{\epsilon_{\text{VBM}}-\epsilon_i}\right]
\notag \\ & \times 
\langle\phi_{\text{CBM}}| \frac{|\phi_i|^2}{\rho_0} |\phi_{\text{CBM}}\rangle,
\label{Eqn:delta_GLLB}
\end{align}
where $\phi_{\text{CBM}}$ is the KS spinorbital corresponding to $\epsilon_{\text{CBM}}$, the conduction band minimum (CBM), $\epsilon_{\text{VBM}}$ is the valence band maximum (VBM), $\rho_0$ is the ground-state density and $K_{\text{xc}}=8\sqrt{2}/(3\pi^2)\approx 0.382$. 
The calculation of the $\Delta^{\text{dd}}$ scales linearly with the system size and, therefore, the numerical speed-up with respect to, for example, a $GW$ calculation is enormous.
Moreover, only ground-state KS quantities enter in Eq.~(\ref{Eqn:delta_GLLB}), it is hence free of convergence problems which can affect the calculation of the $GW$ self-energy. Fundamental gaps calculated using the derivative discontinuity in Eq.~\ref{Eqn:delta_GLLB} have been reported for a large number of 3D and 2D materials.\cite{Kuisma_PRB2010,Baerends_PCCP2017,Thygesen_EESci2012,Thygesen_PRB2013,Thygesen_JPC2015} In general, the results are excellent.

For $f_{\text{xc}}^{(2)}$ we will use the polarization functional (PF)~\cite{Berger} which was designed to take into account excitonic effects in solids.
It is based on a model describing a system with a small dielectric constant and a strongly bound exciton having a large spectral weight.
Nevertheless, we have shown that the PF works well also for systems such as Si and GaAs, which have a large macroscopic dielectric constant and no strongly bound excitons.

The polarization functional is a simple correction to $\chi_e^{\text{RPA}}(\omega)$, the electric susceptibility in the random-phase approximation (RPA), according to~\cite{deboeij,Berger}
\begin{equation}
\left[\chi_e(\omega)\right]^{-1}=\left[\chi_e^{\text{RPA}}(\omega)\right]^{-1}-\alpha,
\label{Eqn:Dyson_alpha}
\end{equation}
where $\chi_e(\omega)$ is the electric susceptibility and $\alpha$ is given by~\cite{Berger,santiago}
\begin{equation}
\alpha=\frac{4\pi}{[\varepsilon^{\text{RPA}}_M(0) - 1]\varepsilon^{\text{RPA}}_M(0)},
\label{Eqn:alphakernel}
\end{equation}
with $\varepsilon^{\text{RPA}}_M(0)$ the RPA macroscopic dielectric function at $\omega=0$. 
From the electric susceptibility the macroscopic dielectric function is obtained as $\epsilon_M(\omega)=1+4\pi\chi_{e}(\omega)$.
The imaginary part of $\epsilon_M(\omega)$ yields the optical absorption spectrum.
We note that the original polarization functional contains an additional dynamical part.~\cite{Berger}
Since this part is mainly important for the description of the Drude-like tail in the absorption spectra of metals~\cite{berger_PRB2006,Ferradas_2018} we will not include it here.
Furthermore, the polarization functional was originally presented in the framework of time-dependent current-density functional theory (TDCDFT)~\cite{DharaGhosh,GhoshDhara,Vignale,Sangalli_2017} in which $\alpha$ is a $3\times 3$ tensor instead of a scalar.
The details of the practical advantages of our TDCDFT approach can be found elsewhere.~\cite{Freddie,Pina,Arjan_2005,Arjan2,Arjan1}

The Pure functional is simply defined as the sum of the two contributions, i.e.,
\begin{equation}
f_{\text{xc}}^{\text{Pure}} = f_{\text{xc}}^{(1,\text{GLLB})} + f_{\text{xc}}^{(2,\text{PF})}.
\end{equation}
It is important to note that the above expression implies that the modified polarizability in Eq.~\eqref{Eqn:modified-chi} has to be used to evaluate $\chi^{\text{RPA}}_e$ in Eqs.~\eqref{Eqn:Dyson_alpha} and \eqref{Eqn:alphakernel}.
Thanks to the simplicity of Eqs.~\eqref{Eqn:modified-chi}-\eqref{Eqn:alphakernel}, the cost of a calculation with the Pure functional equals the cost of a simple RPA calculation.
However, despite the simplicity of the expressions, the Pure functional accurately describes the optical spectra of standard semiconductors and wide-gap insulators.~\cite{PhysRevB.101.115109}

We note that, although $\alpha$ leads to a simple shift of $[\chi_e^{\text{RPA}}(\omega)]^{-1}$, the result is an expression for $\chi_e(\omega)$ which mixes the real and imaginary parts of $\chi_e^{\text{RPA}}(\omega)$ in a non-trival way. 
In particular, the imaginary part of $\chi_e(\omega)$, which is related to the absorption spectrum, becomes

\begin{equation}
\im \chi_e(\omega)=\frac{\im \chi_e^{\text{RPA}}(\omega)}
{\left[1-\alpha\re \chi_e^{\text{RPA}}(\omega)\right]^2+[\alpha\im \chi_e^{\text{RPA}}(\omega)]^2}.
\label{Eqn:exciton}
\end{equation}
It can be readily verified that for $\alpha=0$ this expression reduces to $\chi^{\text{RPA}}_e(\omega)$. 
Excitonic peaks arise when $1-\alpha\re \chi_e^{\text{RPA}}(\omega)=0$ for an energy smaller than the direct gap, since for such an energy $\im \chi_e^{\text{RPA}} = 0$. This equation will prove to be useful in our analysis of the optical spectra of $h$-BN and MoS$_2$ in Section \ref{Results}.

\subsection{The 2D Pure functional}
Given the success of the Pure functional for 3D materials, it would be interesting to see if it can also be successfully applied to 2D materials.
Unfortunately, the PF part of the Pure functional cannot be straightforwardly applied to 2D materials since the macroscopic dielectric function, which appears in Eq.~\eqref{Eqn:alphakernel}, is ill-defined in 2D.\cite{Hueser,Cudazzo,Veniard}
However, following Ref.~\onlinecite{npj2DMaterAppl}, for monolayers we can define the following in-plane ($\varepsilon^{2D}_\parallel$) and out-of-plane ($\varepsilon^{2D}_\perp$) macroscopic dielectric functions, 

\begin{align}
\varepsilon^{2D}_\parallel(\omega)\!&\!=\!1+\left[\varepsilon^{3D}_\parallel(\omega)-1\right]\frac{L_z}{d},\label{Eqn:e_pall_2D}\\
\varepsilon^{2D}_\perp(\omega)\!&\!=\!\left\{1+\left[\frac{1}{\varepsilon^{3D}_\perp(\omega)}-1\right]\frac{L_z}{d}\right\}^{-1},
\label{Eqn:e_pep_2D}
\end{align}
where $\varepsilon^{3D}_\parallel$ and $\varepsilon^{3D}_\perp$ are the in-plane and out-of-plane dielectric functions of a 3D supercell, $L_z$ is the length of the 3D supercell in the direction perpendicular to the monolayer and $d$ the thickness of the monolayer.
The latter is defined as the interlayer distance of the corresponding bulk material.
An expression for $\alpha$ that is equivalent to Eq.~\eqref{Eqn:alphakernel} can then be defined as

\begin{equation}
\alpha=\frac{4\pi}{[\varepsilon^{2D,\text{RPA}}_{\parallel/\perp}(0) - 1]\varepsilon^{2D,\text{RPA}}_{\parallel/\perp}(0)}.
\label{Eqn:alphakernel_2D}
\end{equation}
The 2D Pure functional thus uses Eq.~\eqref{Eqn:alphakernel_2D} for $\alpha$ instead of  Eq.~\eqref{Eqn:alphakernel}.

\begin{table*}[t]
\caption{Calculated and measured 
direct ($E_g^{\text{dir}}$) and indirect band gaps ($E_g^{\text{ind}}$) (in eV) of monolayer and bulk $h$-BN and MoS$_2$. The GLLB-SC values include the derivative discontinuity $\Delta^{dd}$. The DFT values are obtained using LDA for $h$-BN and GGA for MoS$_2$. We note that for MoS$_2$ we report the values that are corrected for relativistic effects. We also report the measured optical gap ($E_g^{\text{opt}}$).
}
    \centering
    %\begin{tabular}{llcccccc}
    \begin{tabular*}{0.70\textwidth}
    {@{\extracolsep{\fill}}lccccccccc}
    \hline
      %&
      & & \multicolumn{4}{c}{$E_g^{\text{dir/ind}}$}
      &$E_g^{\text{opt}}$\\
      \cline{3-6}
      & & DFT & GLLB-SC & $GW$  & Exp 
      & Exp 
         \\
      \hline
      mono $h$-BN 
      &
      & 4.52  
       
      & 7.95 
      & 
      7.25--7.77\footnote{See Refs.~\onlinecite{PhysRevB.94.125303,  PhysRevB.87.235132,  Ferreira:s, 2Ddatabase, PhysRevLett.96.126104}
      }
            & 4.6–7.0 \footnote{See Ref.~\onlinecite{Nagashima}}
      & $>$5.85\footnote{See Ref.~\onlinecite{Stehle}},$\sim$6.17\footnote{See Ref.~\onlinecite{Kun}} 
      \\
      bulk $h$-BN    
      & dir. 
      & 4.44   
          & 7.71  
          & 
          6.28--6.47\footnote{See Refs.~\onlinecite{PhysRevB.98.125206, PhysRevB.97.241114, PhysRevLett.96.026402}}
          & 5.971\footnote{See Ref.~\onlinecite{Watanabe2004}}, 6.4\footnote{See Ref.~\onlinecite{doi:10.1002/pssr.201105190}}
          & 5.822\footnote{See Ref.~\onlinecite{Watanabe2004}} \\
          & ind. & 3.97
      & 7.27 
      &  5.80--5.95\footnote{See Refs.~\onlinecite{PhysRevB.98.125206, PhysRevB.97.241114,  PhysRevLett.96.026402}} 
      & 5.955\footnote{See Ref.\ \onlinecite{2016NaPho..10..262C}}
      \\
          mono MoS$_2$ 
      &
      & 1.62 
      & 2.23  
      & 2.40--2.84\footnote{See Refs.~\onlinecite{PhysRevB.88.045412},
      \onlinecite{Haastrup_2018}, 
      \onlinecite{2Ddatabase},
      \onlinecite{PhysRevB.86.115409}, \onlinecite{PhysRevLett.111.216805}, \onlinecite{PhysRevB.93.235435} and references therein}
      & 2.40--2.5\footnote{See Refs.~\onlinecite{Huang2015, Klots2014}},2.86\footnote{See Ref. \onlinecite{PhysRevB.90.195434}}
      & 1.83--1.92\footnote{See Refs.~\onlinecite{PhysRevLett.105.136805}, 
      \onlinecite{C5NR08219F}, 
      \onlinecite{doi:10.1021/nl302584w}, 
      \onlinecite{Huang2015},\onlinecite{PhysRevB.90.195434}}
      \\
          bulk MoS$_2$   
         & dir. 
         &  1.60 
            & 2.04   
            & 2.07\footnote{See Ref.~\onlinecite{doi:10.1021/jp300079d}}, 2,23\footnote{See Ref.~\onlinecite{PhysRevB.88.045412}}  
            &1.96\footnote{See Ref.~\onlinecite{PhysRevB.86.241201} and references therein}
            &
               1.42\footnote{See Ref.~\onlinecite{PhysRevB.90.195434}},1.88\footnote{See Ref.~\onlinecite{PhysRevB.86.241201} and references therein}
             \\
     & ind. 
        & 1.05 
         & 1.62 
        & 1.23--1.79\footnote{See Ref.~\onlinecite{doi:10.1021/jp300079d,  PhysRevB.86.241201, PhysRevB.88.045412}}
        & 1.23\footnote{See Ref.~\onlinecite{doi:10.1021/j100393a010}}, 1.29\footnote{See       Ref.~\onlinecite{Gmelin}}
        &
        \\
      \hline
    \end{tabular*}
    \label{tab:gap}
\end{table*}

\section{Computational details\label{Comput}}
For our calculations we used two DFT-based codes, namely the Amsterdam Modeling Suite (\textsc{AMS}),\cite{ADF3} to calculate the derivative discontinuity within the GLLB model, and the Vienna Ab-initio Simulation Package (\textsc{Vasp})\cite{Kresse.Furthmuller:1996,Shishkin2006_PRB} with the projector-augmented wave (PAW) method,\cite{Kresse.Joubert:1999:PRB} to calculate the RPA dielectric function. The polarization functional is applied through a post-processing procedure. \footnote{We checked that the dispersion of the bands within GLLB-SC (calculated with ADF) and the dispersion of the bands in LDA/GGA (calculated with \textsc{Vasp}) only differ slightly. For example, by aligning the LDA/GGA and GLLB-SC direct band gap, the error on the indirect band gap is 0.03 eV for bulk $h$-BN and 0.13 eV for bulk MoS$_2$. These differences can be deduced from the gaps reported in Table~\ref{tab:gap}.}

As explained below, we studied both the monolayer and the bulk of $h$-BN and MoS$_2$. For the calculations of the bulk systems we used the following lattice parameters: $a=2.50$ \AA{} and $c=6.66$ \AA{} for bulk $h$-BN, 
$a=3.18$ \AA{} and $c=12.7$ \AA{} for bulk MoS$_2$. For the computation of the monolayer systems
we used a periodic supercell of length $L_z$ in the direction perpendicular to the plane; we used $L_z$ = 25 \AA{} for $h$-BN  and $L_z$ = 20 \AA{} for MoS$_2$. 

The calculations of the derivative discontinuity are done within the GLLB-SC
xc potential,\cite{C7CP02123B, PhysRevB.82.115106, PhysRevA.51.1944} which is based on the PBEsol\cite{PhysRevLett.100.136406} correlation potential and uses the GLLB approximation to the exchange optimized effective potential.
We use the QZ4P (quadruple-$\zeta$+4 polarization functions) basis set for $h$-BN and TZ2P (triple-$\zeta$+2 polarization functions) basis set for MoS$_2$ provided by AMS.

The dielectric functions are calculated on top of an LDA, for $h$-BN, and a GGA-PW91,\cite{PhysRevB.46.6671} for MoS$_2$, ground-state calculation. The LDA/GGA gaps are corrected using the following shift ($\Delta^{\text{shift}}=E_{g}^{\text{dir},\text{GLLB}}-E_{g}^{\text{dir},\text{LDA/GGA}}$, with $E_g^{\text{dir}}$ the direct band gap):
 3.43 eV for monolayer $h$-BN, 
3.27 eV for bulk $h$-BN,
0.61 eV for monolayer MoS$_2$, 
and 0.44 eV for bulk MoS$_2$. 

For MoS$_2$ we also included relativistic effects (scalar and spin-orbit coupling). Since in AMS the spin-orbit correction is not available yet for the GLLB functionals, we have estimated this correction at the level of LDA/GGA.\footnote{We found that spin-orbit coupling  reduces the direct band gap by 0.04 eV in the bulk and by 0.08 eV in the monolayer.} 

For the calculation of the dielectric functions 
we used a $60\times60\times1$ $\vect{k}$ point grid for monolayer $h$-BN, a $60\times60\times4$ $\vect{k}$ point grid
for bulk $h$-BN, a $100\times100\times1$ $\vect{k}$ point grid for monolayer MoS$_2$, and a
$70\times70\times4$ $\vect{k}$ point grid for bulk MoS$_2$.
We used an energy cutoff of 250 eV for both $h$-BN and MoS$_2$.
Local fields are included in all the calculations.
The calculated dielectric functions are broadened with a Lorentzian of 0.1 eV for $h$-BN, and 0.02 eV for MoS$_2$.

\section{Results and discussion\label{Results}}
In this section we discuss our results for the dielectric functions of monolayer $h$-BN and MoS$_2$. 
Furthermore, to get more insights into the performance of our protocol we also calculated the dielectric functions of bulk $h$-BN and MoS$_2$.

Let us first have a look at the band gaps of these materials which are important for the correct description of the absorption onset.
In Tab.~\ref{tab:gap} we report the GLLB-SC values we obtained for the direct and indirect gap for all the systems studied. For comparison we report experimental (when available) as well as $GW$ values, which are, however, quite sensitive to the choice of the truncated Coulomb interaction, the level of self-consistency, and various other convergence parameters.
We observe that the GLLB-SC functional tends to overestimate the gaps with respect to experiment, except for monolayer MoS$_2$. Nevertheless, it seems that the GLLB-SC gaps are in reasonable agreement with the $GW$ gaps except for bulk $h$-BN.

Let us now study the dielectric functions of the monolayers.

\subsection{Monolayer}
\subsubsection{$h$-BN}
In the upper panel of Fig.~\ref{Fig:absorption_monolayer} we report the absorption spectra, i.e., the in-plane component of the imaginary part of the dielectric function, of monolayer $h$-BN calculated using the Pure functional with the dielectric function defined in Eq.~\eqref{Eqn:e_pall_2D}. Unfortunately we have not found experimental spectra for the monolayer (to the best of our knowledge experimental data are available only for $h$-BN on a substrate), therefore we compared our results with BSE results from literature. 
The spectrum obtained with the Pure functional shows two characteristic structures: an intense peak at 7.86 eV, which can be identified as a bound exciton because for this energy the denominator of Eq.~\eqref{Eqn:exciton} vanishes, i.e., 
$1-\alpha\mbox{Re} \chi_e^{\text{RPA}}(\omega=7.86 \text{ eV}) = 0$, and a broad shoulder around 8.5 eV. These structures are in qualitative agreement with the BSE results\cite{PhysRevLett.96.126104} also reported in Fig.~\ref{Fig:absorption_monolayer}. We note, however, that the peak at 6.5 eV in the BSE spectrum is also a bound exciton, contrary to the Pure functional spectrum, where the shoulder can be traced back to the underlying RPA spectrum, i.e., to interband transitions. 
The Pure functional, in its static approximation, can indeed reproduce only one bound exciton. 
The spectrum is blue shifted by about 2 eV with respect to the BSE results. This is caused by the large difference in the exciton binding energies.
The Pure functional predicts an excitonic binding energy of about 0.09 eV,
while the BSE excitonic binding energies reported in literature are in the range of 1.50--2.19 eV (see Ref.\ \onlinecite{Ferreira:19} and references therein). 

Increasing $\alpha$ shifts the main peak towards the BSE main peak but at the cost of a huge overestimation of its intensity as shown in Fig.~\ref{Fig:absorption_bulk&monolayer}. The fact that the intensity of the exciton increases by increasing $\alpha$ is explained in App.~\ref{App:intensity}.  

\subsubsection{MoS$_2$}
\begin{figure}[t]
\begin{center}
\includegraphics[width=0.42\textwidth]{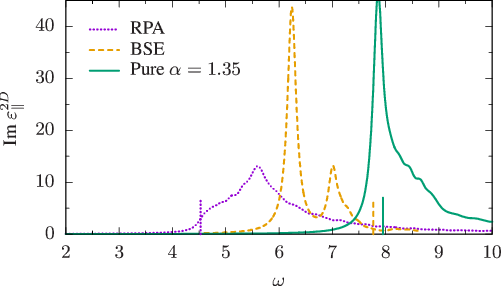}\\\vspace{8pt}
\includegraphics[width=0.42\textwidth]{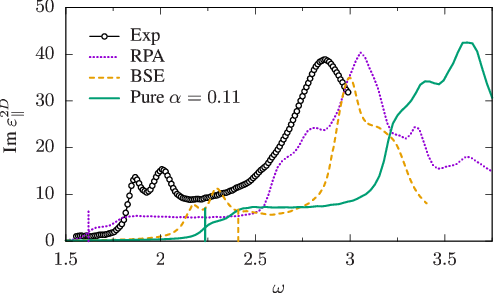}
\caption{Dielectric function of monolayer $h$-BN (upper panel) and MoS$_2$ (lower panel): imaginary part of the in-plane component $\varepsilon_\parallel$ within RPA (violet dotted line), BSE\cite{PhysRevLett.96.126104, PhysRevB.88.045412} (orange dashed line), and Pure functional (green solid line). For MoS$_2$ we also report the experimental spectrum.\cite{PhysRevB.90.205422} 
Vertical lines indicate the position of $E_g^{\text{dir,LDA/GGA}}$ (violet dotted line), $E_g^{\text{dir},GW}$ (orange dashed line), and $E_g^{\text{dir,GLLB}}$ (green solid line).
}
%\vspace{2.5mm}
\label{Fig:absorption_monolayer}
\end{center}
\end{figure}

\begin{figure*}[t]
\begin{center}
\includegraphics[width=0.24\textwidth]{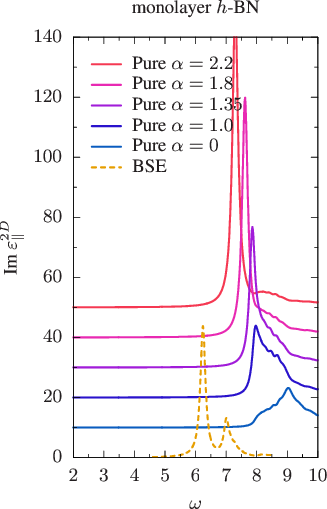}
\includegraphics[width=0.24\textwidth]{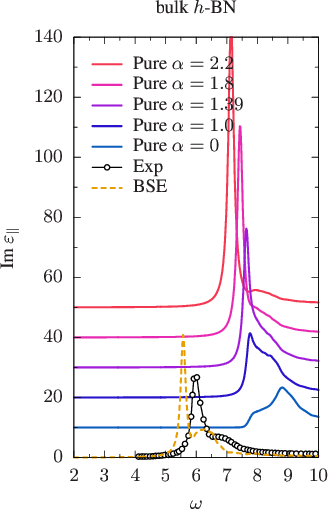}
\includegraphics[width=0.24\textwidth]{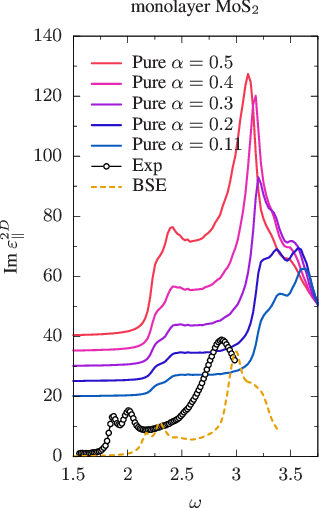}
\includegraphics[width=0.24\textwidth]{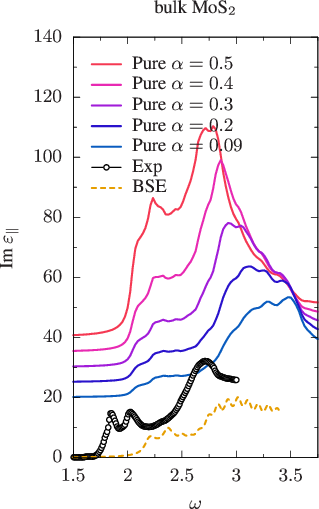}
\caption{Dielectric function of (from left to right) monolayer $h$-BN, bulk $h$-BN, monolayer MoS$_2$ and bulk MoS$_2$: imaginary part of the in-plane component $\varepsilon_\parallel$ within the Pure functional (solid lines of different colors according to the value of $\alpha$),
within BSE\cite{PhysRevLett.96.126104, PhysRevB.97.241114, nomadBSE, PhysRevB.88.045412} (orange dashed line), and from experiments\cite{PhysRevB.90.205422, PhysRevB.40.7852, Beal_1979} (circles). The Pure functional curves are vertically shifted for clarity.}
\label{Fig:absorption_bulk&monolayer}
\end{center}
\end{figure*}

In the lower panel of Fig.~\ref{Fig:absorption_monolayer} we report the absorption spectrum, i.e., the in-plane component of the imaginary part of the dielectric function, of monolayer MoS$_2$ calculated using the Pure functional, with the dielectric function defined in Eq.~\eqref{Eqn:e_pall_2D}. 
We compare our results with those obtained in experiment\cite{PhysRevB.90.205422} and with BSE results.\cite{PhysRevB.88.045412}$^,$\footnote{We note that for MoS$_2$ (both monolayer and bulk) the BSE spectra found in literature are presented in arbitrary units; for a comparison with our spectra we have hence rescaled them in such a way that the corresponding RPA spectra are roughly on top of ours.} The experimental spectrum shows a double-peak structure at low energy (specifically at 1.86 eV and 2.01 eV), which is also visible in the BSE spectrum, although blue-shifted by $\approx$ 0.3 eV. This double-peak structure appears as a double shoulder in the spectrum calculated using the Pure functional, which is blue shifted by 0.07 eV with respect to the BSE spectrum, and by 0.38 eV with respect to experiment. Moreover the excitonic binding energy calculated by the Pure functional is zero as in the RPA, while the BSE predicts a value of 0.24 eV,\cite{PhysRevB.88.045412}, and experiment give a value of 0.48 eV. \cite{Huang2015}
This is caused by an underestimation of $\alpha$, for which the Pure functional yields 0.11, whereas one would need a much larger value for the exciton to appear.  In Fig.~\ref{Fig:absorption_bulk&monolayer} we show how the absorption spectrum changes by increasing $\alpha$.
We note that the peak that appears for $\alpha=0.5$ is not yet a bound exciton, i.e., a solution of $1-\alpha \re \chi_e^{\text{RPA}}(\omega)=0$. To observe a bound exciton as in $h$-BN one would need an even larger $\alpha$ ($\alpha>0.7$), which, however, could deform the rest of the spectrum.

In conclusion our TDDFT protocol can describe the qualitative features of the optical spectrum of monolayer $h$-BN, although the spectrum is blue-shifted and the binding energy is underestimated with respect to BSE results from the literature. Unfortunately there are no experimental data to compare with. Similar trends are observed for monolayer MoS$_2$ when compared to BSE as well as to experimental spectra; for this system, however, the double-peak exciton is reproduced only as a double shoulder.

\subsection{Bulk}
To better understand the results we obtained for the monolayers, and, in particular, to distinguish between the role played by the Pure functional and the role played by the 2D macroscopic dielectric functions in Eqs.~\eqref{Eqn:e_pall_2D}-\eqref{Eqn:e_pep_2D}, we studied the optical response of the corresponding bulk materials as well. 
In this case we can use the standard definition of the macroscopic dielectric function and test, hence, the performance of the Pure functional.
\subsubsection{$h$-BN}
\begin{figure*}[t]
\begin{center}
\includegraphics[width=0.42\textwidth]{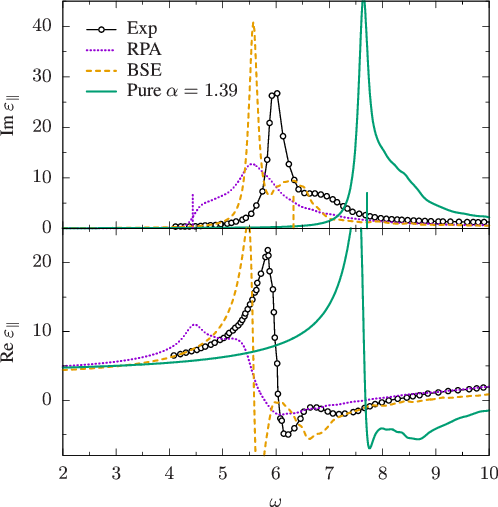}\hspace{15pt}
\includegraphics[width=0.42\textwidth]{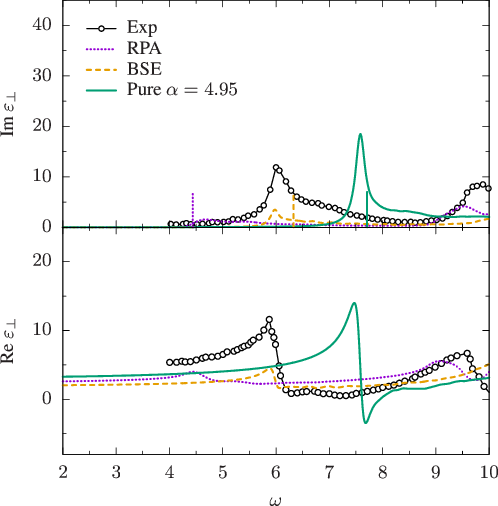}
\caption{Dielectric function of bulk $h$-BN: real and imaginary parts of in-plane component $\varepsilon_\parallel$ (left panel) and of the out-of-plane component $\varepsilon_\perp$ (right panel) within RPA (dotted violet line), BSE\cite{PhysRevB.97.241114, nomadBSE} (dashed orange line) and Pure functional (solid green line) compared to experiment\cite{PhysRevB.40.7852} (circles). A Lorentzian broadening of 0.1 eV is applied to all calculated spectra. Vertical lines indicate the position of $E_g^{\text{dir,LDA/GGA}}$ (violet dotted line), $E_g^{\text{dir},GW}$ (orange dashed line), and $E_g^{\text{dir,GLLB}}$ (green solid line).
}
%\vspace{2.5mm}
\label{Fig:absorption$h$-BNbulk}
\end{center}
\end{figure*}

In Fig.~\ref{Fig:absorption$h$-BNbulk} we report the in-plane ($\varepsilon_\parallel$) and out-of-plane ($\varepsilon_\perp$) components of the real and imaginary parts of the dielectric function of bulk $h$-BN. Our results obtained with the Pure functional are compared to the RPA spectra, the BSE spectra reported in Refs.~\onlinecite{PhysRevB.97.241114} and \onlinecite{nomadBSE}, and to experiment.\cite{PhysRevB.40.7852}  As for the monolayer, bulk $h$-BN exhibits a strongly bound exciton.
This exciton, which appear in the experimental spectrum around 6 eV, is completely absent in the RPA spectrum.
Except an overall blue shift of about 1.6 eV, the Pure functional gives a spectrum in qualitative agreement with the experimental and the BSE spectrum, with a bound exciton at 7.64 eV and a broad shoulder at about 8.4 eV. The blue shift is due to the overestimation of the direct band gap by the GLLB-SC model as can be seen in Table \ref{tab:gap}, but in part also to the PF. Overall, when comparing our spectra to those obtained within the BSE, we observe a similar trend as for monolayer $h$-BN.
This seems to validate the expression for the 2D macroscopic dielectric function defined in Eqs.~\eqref{Eqn:e_pall_2D}-\eqref{Eqn:e_pep_2D}.

The exciton binding energy predicted by the Pure functional is 0.07  eV, which is about half the binding energy found in experiment, which is 0.149 eV.\cite{Watanabe2004}
The BSE binding energy is 0.75 eV,\cite{PhysRevB.97.241114, nomadBSE} largely overestimating the experimental value.A similar trend is found also for the out-of-plane component of the dielectric function, with a slightly better description of the binding energy (0.12 eV from the Pure functional). By increasing the value of $\alpha$, the main excitonic peak becomes much more prominent and moves to lower energies and the binding energy improves, as shown in Fig.~\ref{Fig:absorption_bulk&monolayer}, where we reported the in-plane component $\varepsilon_\parallel$ (a similar trend occurs for the out-of-plane component as well). We notice that the optimal value of $\alpha$ to get the experimental binding energy is $\approx$1.57, which is rather close to the value of 1.39 calculated with the Pure functional.

\begin{table*}[t]
    \centering
    \caption{Calculated and measured macroscopic dielectric constant [$\mbox{Re} \varepsilon_{\parallel/\perp}(\omega=0)$]}
    \begin{tabular*}{0.70\textwidth}
    {@{\extracolsep{\fill}}lrccccc}
    %{lrccccc}
    \hline
       & & RPA@LDA/GGA& RPA@LDA/GGA+$\Delta^{\text{shift}}$  & Pure 
        & Exp\\
       \hline
      bulk $h$-BN & $\varepsilon_\parallel$ & 4.60& 3.55 
       & 4.55  
       & 4.95
       \footnote{See Refs.~\onlinecite{PhysRev.146.543, PhysRevB.40.7852}}\\
      & $\varepsilon_\perp$ & 2.52 &  2.17  & 3.17 
      & 
      4.10\footnotemark[1] \\
      bulk MoS$_2$ & $\varepsilon_\parallel$ 
      & 13.42& 12.10  
      & 13.10 
      & 15--16\footnote{See Ref.~\onlinecite{Beal_1979}}\\
      & $\varepsilon_\perp$ & 4.84&  4.59   
      & 5.59  
      & 
      \\
      \hline
    \end{tabular*}
    \label{tab:dielconst}
\end{table*}

Since the Pure functional is sensitive to the RPA macroscopic dielectric constant (see Eq.~\eqref{Eqn:alphakernel})
we report them in Tab.~\ref{tab:dielconst} for the materials we study here using various methods.
The RPA dielectric constants calculated on top of a GLLB-SC corrected LDA/GGA band structure (RPA@LDA/GGA+$\Delta^{\text{shift}}$ in Table \ref{tab:dielconst}) are systematically smaller than those calculated on top of a LDA/GGA band structure (RPA@LDA/GGA). 
We note that, by definition, $\varepsilon_M^{\text{Pure}}(0) = \varepsilon_M^{\text{RPA}}(0) + 1$, which can be verified from Eqs.~\eqref{Eqn:Dyson_alpha} and \eqref{Eqn:alphakernel}.
This increase of the static dielectric constant improves the agreement with experiment, although they are still smaller. 
The important point is that the dielectric constant for bulk $h$-BN is relatively small; together with the fact that the spectrum is dominated by a bound exciton, $h$-BN is similar to the model the polarization functional was derived from. 
This might explain why the spectra calculated using the Pure functional are in relatively good agreement with the experiments for bulk $h$-BN. 

\begin{figure*}[t]
\begin{center}
\includegraphics[width=0.42\textwidth]{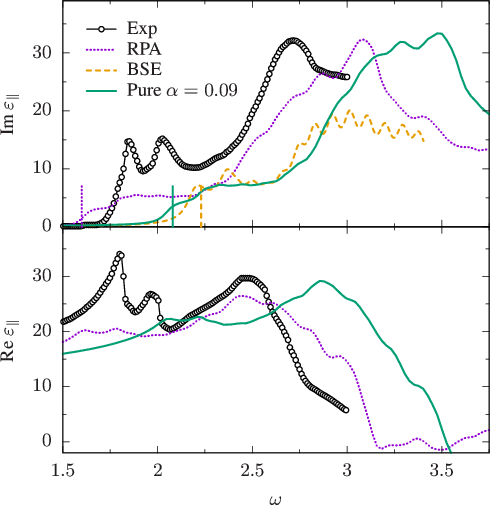}\hspace{15pt}
\includegraphics[width=0.42\textwidth]{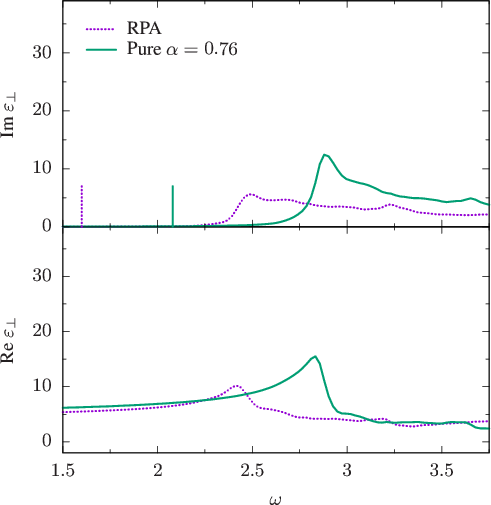}
\caption{
Dielectric function of bulk MoS$_2$: real and imaginary parts of the in-plane component $\varepsilon_\parallel$ (left panel) and of the out-of-plane component $\varepsilon_\perp$ (right panel) within RPA (violet dotted line), BSE\cite{PhysRevB.88.045412} (orange dashed line) and Pure functional (green solid line) compared to experiment\cite{Beal_1979} (circles). 
A Lorentzian broadening of 0.02 eV and 0.05 eV is applied to the Pure functional and BSE spectra, respectively.
Vertical lines indicate the position of $E_g^{\text{dir,LDA/GGA}}$ (violet dotted line), $E_g^{\text{dir},GW}$ (orange dashed line), and $E_g^{\text{dir,GLLB}}$ (green solid line).
}
\label{Fig:absorptionMoS2bulk}
\end{center}
\end{figure*}

\subsubsection{MoS$_2$}
The dielectric function of bulk MoS$_2$ is reported in Fig.~\ref{Fig:absorptionMoS2bulk}. Unfortunately we were not able to find any experimental nor BSE data for the  out-of plane component ($\varepsilon_\perp$), but we reported, nevertheless, our results for completeness.  
Also in this case we find a similar trend as for the monolayer.
The value of $\alpha$ calculated using Eq.~\eqref{Eqn:alphakernel} is too small ($\alpha=0.09$) due to the large value of the macroscopic dielectric constant (see Tab.~\ref{tab:dielconst}). This case is beyond the model employed to derive the PF. In this case $1-\alpha\re \chi_e^{\text{RPA}}(\omega)=0$ in the denominator of Eq.~\eqref{Eqn:exciton} has no solution and the various terms in Eq.~\eqref{Eqn:exciton} simply rescale the RPA spectrum (See Fig.~\ref{Fig:absorptionMoS2bulk}). Indeed the spectrum calculated using the Pure functional is similar to the RPA spectrum (except for the larger absorption onset).
As for the monolayer the two peaks in the experimental spectrum\cite{Beal_1979} at 1.84 eV and 2.02 eV are reproduced as shoulders by the Pure functional. 
Moreover the two structures are blue shifted by about 0.25 eV with respect to experiment. 
In the BSE spectrum the two structures are more evident but also blue shifted, by about 0.35 eV. The experimental binding energy of 0.08-0.09 eV\cite{PhysRevB.11.905,PhysRevB.86.241201} is underestimated by both the BSE, with a binding energy of 0.03 eV \cite{PhysRevB.88.045412} corresponding to the BSE spectrum reported in Fig.~\ref{Fig:absorptionMoS2bulk} (although other values reported in literature overestimate experiment, see e.g.\ Ref.~\onlinecite{PhysRevB.86.241201}) and the Pure functional, with a binding energy of 0 eV.  
Finally we note that a low-frequency peak emerges in the Pure functional spectrum by increasing $\alpha$. As shown in Fig.~\ref{Fig:absorption_bulk&monolayer}, a peak becomes quite visible at $\alpha=0.5$, but, as for the monolayer, this is not yet a bound exciton. 

In conclusion our TDDFT protocol can describe the qualitative features of the optical spectrum of bulk $h$-BN, although the spectrum is blue-shifted and the binding energy is halved with respect to experiment. However, the computationally more expensive BSE produces a blue-shifted spectrum too and moreover largely overestimates the binding energy. The situation is exacerbated in bulk MoS$_2$, where, moreover, as in the monolayer, the two lowest-energy peaks in the experimental spectrum are reproduced as shoulders by our protocol. However, also the BSE produces a blue-shifted spectrum and underestimates the binding energy. 

\section{Conclusions and Perspectives\label{Conclusions}}
We explored the description of the optical response of 2D monolayers within a pure density-based approach using the Pure functional. 
Our protocol, originally devised for bulk materials, is generalized by using a well-behaved 2D macroscopic dielectric function.
When applied to $h$-BN monolayer, which shows a strong bound exciton, this approach can well reproduce the optical spectra, except for a rigid blue shift with respect to the BSE results (unfortunately experimental data are not available). 
Instead, in MoS$_2$ monolayer, the excitonic peaks are only described as shoulders. 
We found a similar trend also for the corresponding bulk systems of $h$-BN and MoS$_2$. 
This validates the 2D macroscopic dielectric function we use in our approach and traces back the observed trends to the Pure functional.

The Pure functional has been designed for isotropic systems with a small dielectric constant.
While $h$-BN has a relatively small macroscopic dielectric constant (an experimental value of 4.95 for the in-plane component in the bulk), MoS$_2$ has a rather large dielectric constant (an experimental value of 15--16 for the in-plane component in the bulk).
While we have found that the Pure functional yields accurate spectra also for 3D isotropic systems with a large macroscopic dielectric constant such as Si and GaAs, it seems that for layered materials it is more sensitive to the value of the macroscopic dielectric constant.
This suggests that the dependence of $\alpha$ on $\epsilon_M(0)$ has to be modified and/or that the macroscopic dielectric function does not provide enough information about the screening of the electron-hole interaction in layered systems and one should include information from the microscopic dielectric function.

\section*{Conflicts of interest}
There are no conflicts of interest.

\section*{Acknowledgements}
This work has been supported through the EUR grant NanoX ANR-17-EURE-0009 in the framework of the ``Programme des Investissements d'Avenir'' and by ANR (project no.\ ANR-18-CE30-0025). 
This work was granted access to the HPC resources of CALMIP supercomputing center under the allocation 2020-20036. The authors are grateful to Davide Sangalli and Alejandro Molina-S\'anchez for useful discussions about the BSE calculations.

\appendix   
\section{Dependence of the excitonic intensity on $\alpha$}\label{App:intensity}
We start from Eq.~(\ref{Eqn:exciton}). In the vicinity of $\omega_{be}$ we can approximate $\im \chi_e(\omega)$ as
\begin{eqnarray}
\im \chi_e(\omega)&\approx&\frac{\im \chi_e^{\text{RPA}}(\omega)}
{\left[\alpha B (\omega-\omega_{be})\right]^2+[\alpha\im \chi_e^{\text{RPA}}(\omega)]^2}\nonumber\\
&=&\frac{1}{\alpha^2 |B|}\frac{\eta}
{(\omega-\omega_{be})^2+\eta^2},\nonumber
\label{Eqn:exciton_2}
\end{eqnarray}
where we Taylor expanded up to first order $\re \chi_e^{\text{RPA}}(\omega)$ and we used the condition $1/\alpha=\re \chi_e^{\text{RPA}}(\omega_{be})$, necessary to get an exciton. Here $B=\frac{d \re \chi_e^{\text{RPA}}(\omega)}{d \omega}|_{\omega=\omega_{be}}$ and $\eta=\im \chi_e^{\text{RPA}}(\omega)/|B|$.

Since $\im \chi_e^{\text{RPA}}(\omega)=0$ in the vicinity of $\omega_{be}$, we can consider the limit $\eta\rightarrow 0^+$. We get
\begin{equation}
\im \chi_e(\omega)=\frac{\pi}{\alpha^2 |B|}\delta(\omega-\omega_{be}).\nonumber
\end{equation}

To get an estimate of the dependency of the excitonic intensity on $\alpha$, we should make explicit the dependence of $B$ on $\alpha$. We assume that $\re \chi^{RPA}_e(\omega)$ has the following form for $\omega<E_{g}^{\text{dir}}$: 
$$\re \chi^{RPA}_e(\omega)=\chi^{\infty}+\frac{A}{\omega-\bar{\omega}},$$
with $A<0$ and $\chi^{\infty}=\re \chi^{RPA}_e(\omega\rightarrow\infty)>0$. 

Since to get an exciton the condition $1/\alpha=\re \chi^{RPA}_e(\omega_{be})$ has to be satisfied, we then get
$$\re \chi^{RPA}_e(\omega_{be})=\chi^{\infty}+\frac{A}{\omega_{be}-\bar{\omega}}=1/\alpha$$
from which
$$\omega_{be}=\frac{A}{1/\alpha-\chi^{\infty}}+\bar{\omega}$$
and
$$B=-\frac{A}{(\omega_{be}-\bar{\omega})^2}
=-\frac{(1-\alpha\chi^{\infty})^2}{\alpha^2A}$$
The intensity of the excitonic peak is thus
$$\frac{\pi}{\alpha^2|B|}=
\frac{\pi|A|}{(\alpha\chi^{\infty}-1)^2}.$$

We note that a necessary condition to have $\omega_{be}>0$ is
$1/\alpha>\re \chi_e^{RPA}(0)=\chi^{\infty}-A/\bar{\omega}>\chi^{\infty}$, where we have used the fact that $A<0$ and $\bar{\omega}>0$.
It can be readily verified that in the domain $\alpha_{\text{min}}<\alpha<1/\re \chi_e^{RPA}(0)$, with $\alpha_{\text{min}}>0$ the minimum value of $\alpha$ to get an exciton, the excitonic intensity increases monotonically with respect to $\alpha$.

%\bibliography{notes.bib}

\end{document}